\documentclass[twocolumn]{aastex631}
\usepackage{amsmath,amssymb}
\usepackage{mathtools}
\usepackage{tabularx}

\newcommand{\Msun}{M_{\odot}}

\newcommand{\Ye}{Y_{\rm e}}

\shorttitle{Linking Analytic Light Curve Models to Physical Properties of Kilonovae}
\shortauthors{Kitamura et al.}
\graphicspath{{./}{figures/}}

\begin{document}

\title{Linking Analytic Light Curve Models to Physical Properties of Kilonovae}

\author{Ayari Kitamura}
\affiliation{Astronomical Institute, Tohoku University, Aoba, Sendai 980-8578, Japan}

\author[0000-0003-4443-6984]{Kyohei Kawaguchi}
\affiliation{Max-Planck-Institut f\"ur Gravitationsphysik (Albert-Einstein-Institut), Am M\"uhlenberg 1, D-14476 Potsdam-Golm, Germany}
\affiliation{Institute for Cosmic Ray Research, The University of Tokyo, Kashiwa, Chiba 277-8582, Japan}
\affiliation{Center for Gravitational Physics, Yukawa Institute for Theoretical Physics, Kyoto University, Kyoto 606-8502, Japan}

\author[0000-0001-8253-6850]{Masaomi Tanaka}
\affiliation{Astronomical Institute, Tohoku University, Aoba, Sendai 980-8578, Japan}
\affiliation{Division for the Establishment of Frontier Sciences, Organization for Advanced Studies, Tohoku University, Sendai 980-8577, Japan}

\author[0000-0001-6467-4969]{Sho Fujibayashi}
\affiliation{Frontier Research Institute for Interdisciplinary Sciences, Tohoku University, Sendai 980-8578, Japan}
\affiliation{Astronomical Institute, Tohoku University, Aoba, Sendai 980-8578, Japan}
\affiliation{Max-Planck-Institut f\"ur Gravitationsphysik (Albert-Einstein-Institut), Am M\"uhlenberg 1, D-14476 Potsdam-Golm, Germany}



\begin{abstract}

In binary neutron star mergers, lanthanide-rich dynamical ejecta and lanthanide-poor post-merger ejecta have been often linked to the red and blue kilonova emission, respectively.
However, analytic light curve modeling of kilonova often results in the ejecta parameters that are at odds with such expectations.
To investigate the physical meaning of the derived parameters, we perform analytic modeling of the kilonova light curves
calculated with realistic multi-dimensional radiative transfer based on the numerical relativity simulations. 
Our fiducial simulations adopt a faster-moving, less massive dynamical ejecta and slower-moving, more massive post-merger ejecta.
The results of analytic modeling, however, show that the inferred ``red'' component is more massive and slower, while the ``blue'' component is less massive and faster, 
as also inferred for GW170817/AT2017gfo.
This suggests that the parameters derived from light curve modeling with an analytic model do not represent the true configuration of the kilonova ejecta.
We demonstrate that the post-merger ejecta contributes to both blue and red emissions:
the emission from the post-merger ejecta is absorbed and reprocessed to red emission by the dynamical ejecta with a higher lanthanide fraction.
Our results caution against separately discussing the origins of red and blue components derived from the analytic models.
Despite of the challenges in the parameter estimation, 
we show that the estimate of the total ejecta mass is rather robust within a factor of a few, reflecting the total luminosity output.
To derive the reliable total ejecta mass, multi-epoch observations in near-infrared wavelengths near their light curve peaks are important.

\end{abstract}



\section{Introduction} 
\label{sec:intro}

\begin{figure}
  \begin{center}
    \includegraphics[width=8cm]{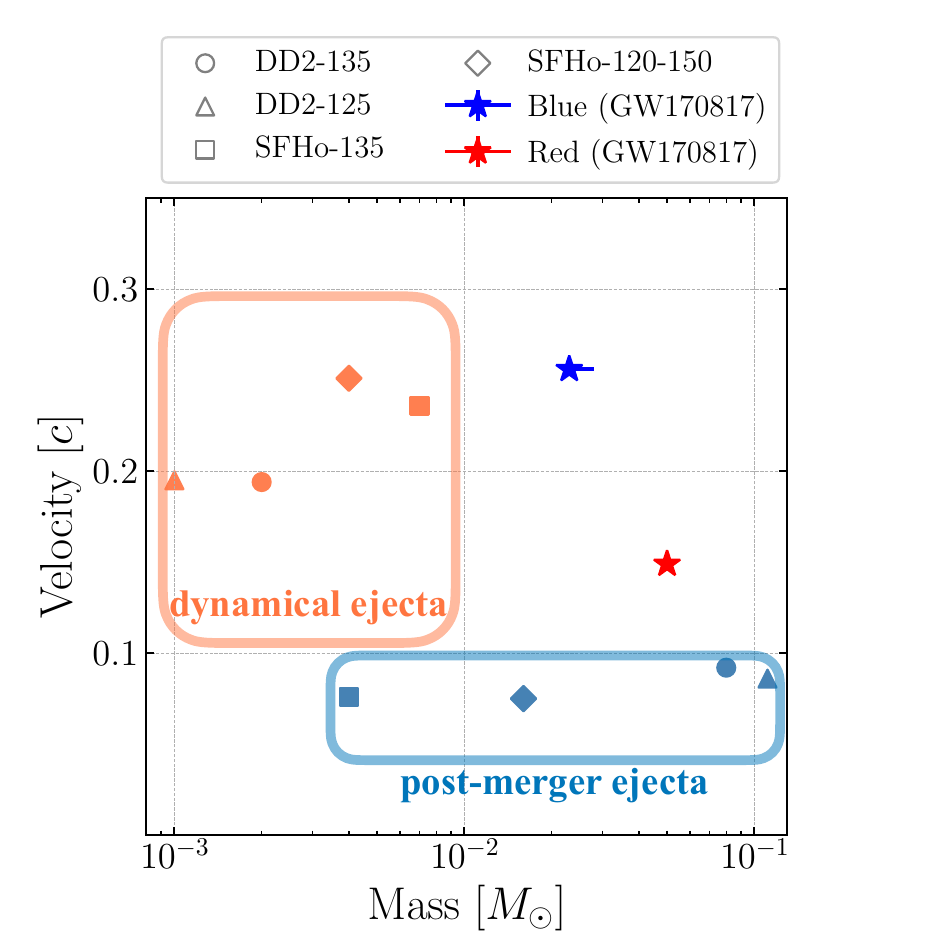}
    \caption{Relationship between mass and velocity of the ejecta from neutron star mergers.
    The blue and red points represent the blue and red components from the analytic model fitting of GW170817/AT2017gfo \citep{villar17}. Other symbols represent four different NR simulations: blue and red colors indicate the post-merger and dynamical ejecta, respectively \citep{Fujibayashi20c, Fujibayashi23}.
  }
\label{fig:obs_Mv}
\end{center}
\end{figure}

The merger of neutron stars is considered as a prime candidate for the origin of $r$-process elements \citep{lattimer74, eichler89, cowan21}. During the mergers, a small fraction of mass is ejected \citep{rosswog99, Hotokezaka13}, synthesizing various radioactive nuclei \citep{freiburghaus99}. Then, decays of the newly synthesized nuclei power electromagnetic wave emissions especially in the optical and near-infrared (NIR) wavelengths, which is called kilonova (\citealt{Li_1998, kulkarni05, metzger10}, see \citealt{tanaka16, Fernandez&Metzger16, metzger17} for reviews). 
Since the kilonova light curves are mainly characterized by ejecta mass ($M$), velocity ($v$), and opacity ($\kappa$) of the ejecta \citep{Li_1998, Kasen13, Kasen15, barnes16, tanaka17, tanaka18}, we can extract the properties of the ejected material from the analysis of the light curves.

Early radiative transfer simulations suggest that the kilonova light curves can have broadly two components: ``red'' (mainly NIR) emission from the lanthanide-rich ejecta with a high opacity \citep{Tanaka_2013, Kasen13}, and ``blue'' (mainly optical) emission from ejecta, which contains lighter $r$-process elements with a low opacity \citep{metzger14, Kasen15, tanaka18}.
These two components are also naturally predicted from numerical relativity (NR) simulations of the merger (see e.g., \citealt{Shibata17, Perego17, Radice18, Shibata&Hotokezaka19}): fast dynamical ejecta are typically neutron-rich including a low electron fraction material ($Y_{\rm{e}} \lesssim 0.25$, lanthanide-rich), 
while slower post-merger ejecta
have a higher $\Ye$ material ($Y_{\rm{e}} \sim 0.25 - 0.5$, lanthanide-poor).

The first joint observation of gravitational and electromagnetic emission from a neutron star merger was achieved for GW170817/AT2017gfo \citep{abbott17a}. 
The optical and NIR light curves of AT2017gfo show a blue emission at the initial phase, followed by a prolonged red emission at later phases \citep{andreoni17, arcavi17, coulter17, Cowperthwaite17, Diaz17, Drout17, Evans17, Kasliwal17, Lipunov17, Pian17, Shappee17, Smartt17, Tanvir17, Troja17, Utsumi17, Valenti17}.
Since the presence of both blue and red components itself is consistent with theoretical expectations as described above,
analytic modeling of AT2017gfo with a two-component (blue and red) model have been often performed (e.g., \citealt{Cowperthwaite17, Drout17, villar17}). 
A similar two-component model has also been applied for kilonova candidates associated with gamma-ray bursts (e.g., \citealt{O'Connor21, Rastinejad22, Rastinejad24, Yang24}).

However, such two-component analytic modeling of the observed light curve yields the ejecta parameters that are at odds with theoretical prediction of NR simulations.
Figure \ref{fig:obs_Mv} shows the relationship between ejecta mass and velocity derived from NR simulations \citep{Fujibayashi20c, Fujibayashi23} and those estimated from analytic modeling \citep{villar17}. 
There is a clear discrepancy between the NR simulations and estimates from the observations: the dynamical ejecta are often thought to correspond to the red component, but the inferred red component has a higher mass and lower velocity.
Similarly, the inferred blue component has a higher velocity than that of the post-merger ejecta. 
Such a discrepancy has prompted some discussion involving the existence of a physically distinct fast blue component (or fast blue ejecta, e.g.,\citealt{Cowperthwaite17, Kasen17, villar17}).

The discrepancy in the derived parameters may stem from the simplification made in the analytic light curve modeling. 
For example, in typical analytic models, the detailed elemental abundances are not considered and the effects of the abundances are represented by a choice of parameterized constant grey opacity \citep{metzger17}.
In reality, however, the opacities depend on wavelength and time, reflecting the bound-bound transitions of heavy elements \citep{Kasen13, tanaka&hotokezaka13, tanaka20}.
Furthermore, the photons in multiple ejecta components should experience complex radiative transfer processes, including absorption and reprocessing \citep{kawaguchi18, kawaguchi20, kawaguchi21}.
In fact, recent, more realistic end-to-end simulations covering the entire process from the merger to observational outputs
\citep{kawaguchi21, kawaguchi22, kawaguchi23, Just23, Shingles23}
show that the light curves similar to those observed in GW170817/AT2017gfo can be realized by the ejecta configuration which NR simulations predict.

Given these situations, 
some studies have attempted to estimate the ejecta parameters using a ``surrogate model'' \citep[e.g.,][]{Ristic:2021ksz,Heinzel:2020qlt}. In this approach, detailed radiative transfer calculations are conducted across parameter grids and then interpolated to enable the fast parameter estimation. 
Some of these studies yield parameter estimates consistent with the NR simulations \citep{almualla21, pang23}. However, some cases result in the estimates similar to those from analytic models \citep{coughlin18, ristic22}, and the estimated parameters are sensitive to the assumptions in the radiative transfer simulations such as ejecta morphology  \citep{almualla21}.
Since realistic radiative transfer simulations based on numerical relativity simulations are still limited, the parameter space that can be used for the parameter estimation is also limited. Thus, it is still challenging to derive physically motivated parameters only with currently available surrogate models.

On the other hand, use of the analytic models has clear advantage in terms of its simplicity. As the analytic model has less degrees of freedom, and hence, less model parameters due to the stronger assumptions, the results are less affected by the details of the model compared to the surrogate models.
Also, it is straightforward to explore the wide parameter range due to the low computation costs.

However,the stronger assumptions can induce large systematic uncertainty in the modeling, and the physical meaning of the derived parameters is not necessarily clear. Hence, to make full use of the analytic models, it is important to understand the physical meaning of the derived parameters.
In this paper, to address this issue, we use realistic, multi-dimensional radiative transfer simulations as mock observational data, for which the actual configuration of the ejecta is known. 
We perform analytic modeling of the simulated light curves and discuss the physical meaning of inferred parameters. 
Then, we extend the modeling to light curves from different simulation models to study the dependence on viewing angles and merger models.

This paper is organized as follows. In Section 2, we describe the method employed in this study, including the radiative transfer simulations that provide mock observational data. 
In section 3, we present the results of analytic modeling of the simulated light curves.
In Section 4, we discuss the interpretation of the inferred parameters. We also explore how missing observational data affects the estimated parameters. Finally, we give a summary of this paper in Section 5.


\section{Methods}
\label{sec:method}

\subsection{Radiative transfer simulation}
\label{subsec:RTsim}

\begin{table*}[th]
  \begin{center}
  \caption{Summary of key model parameters. The columns describe the model name, the adopted EoS, the masses of the neutron stars, ejecta mass and thier average velocity of dynamical and post-merger ejecta, respectively. The dynamical ejecta for SFHo models are defined as the ejecta component already outside the extraction radius ($8\times 10^8\ \rm{cm}$) at $t=0.7\ \rm{s}$, when the $dM/dt$ is at minima (see \citealt{Fujibayashi23}). On the other hand, those for DD2 models are defined as the ejecta component already existed at the beginning of the 2D simulation.  The mass of post-merger ejecta ($M_{\rm{ej}}^{\rm{pm}}$) is determined by subtracting the mass of dynamical ejecta ($M_{\rm{ej}}^{\rm{dyn}}$) from the total ejecta mass evaluated in the NR simulations. The average velocity of the ejecta is calculated with the mass and kinetic energy of the ejecta, following the definition of Eq. (A9) with an ejecta criterion Eq. (4) in \cite{Fujibayashi23}, specifically.}
  \label{tab:model}
  \hspace{-2cm}
    \begin{tabular}{ccccccc}
      \hline \hline
      Model & EOS & $(m1\ [\Msun],m2\ [\Msun])$ & $M_{\rm{ej}}^{\rm{dyn}}\ [10^{-2}M_\odot]$ & $v_{\rm{ej}}^{\rm{dyn}}\ [c]$ & $M_{\rm{ej}}^{\rm{pm}}\ [10^{-2}M_\odot]$ & $v_{\rm{ej}}^{\rm{pm}}\ [c]$\\ \hline
      DD2-135    & DD2 & (1.35, 1.35) & 0.15 & 0.19 & 8.0 & 0.092\\
      DD2-125    & DD2 & (1.25, 1.25) & 0.087 & 0.20 & 11 & 0.086\\
      SFHo135-135& SFHo & (1.35, 1.35) & 0.71 & 0.24 & 0.36 & 0.076\\
      SFHo120-150& SFHo & (1.20, 1.50) & 0.38 & 0.25 & 1.6 & 0.075\\
      \hline
    \end{tabular}
  \end{center}
\end{table*}

\begin{figure*}[htbp]
  \begin{center}
    \includegraphics[width=13cm]{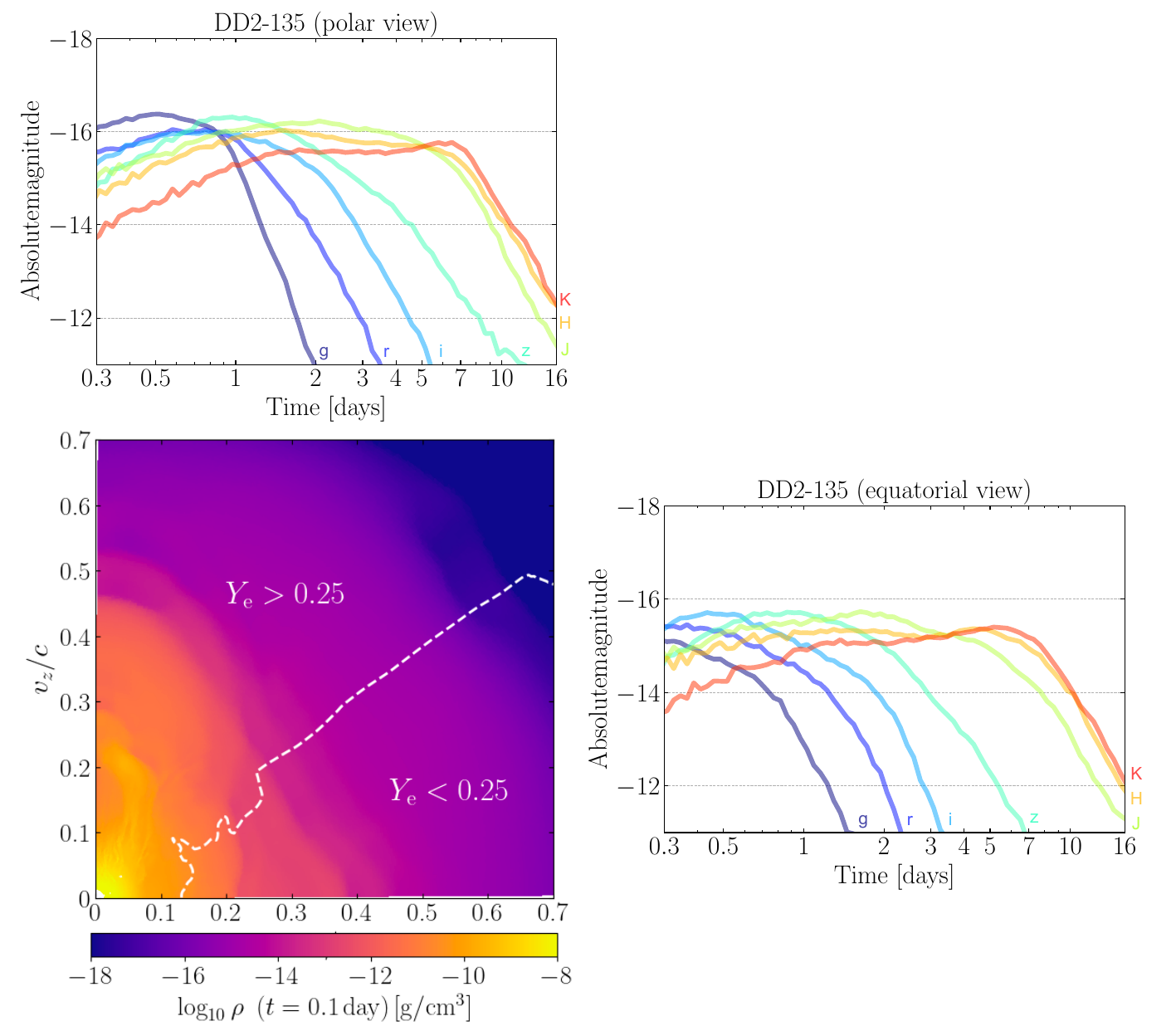}
    \caption{Rest-mass density profile of the ejecta in the meridional plane (the $z$-axis denotes the polar axis) obtained by the hydrodynamics simulation at $t \approx 0.1$ days for DD2-135, along with the multiband light curves resulting from the radiative transfer simulation, observed from the polar angle ($0^{\circ} < \theta < 20^{\circ}$) and the equatorial angle ($86^\circ < \theta < 90^\circ$), respectively \citep{kawaguchi22}.
  }
\label{fig:sim-lc}
\end{center}
\end{figure*}

In this work, we employ the light curves calculated by~\cite{kawaguchi21, kawaguchi22, kawaguchi23} as the inputs for the parameter estimation. These light curves are obtained by performing radiative transfer simulations based on NR simulations of binary neutron stars. The simulations are first performed in 3D to study the inspiral to the merger phase of binary neutron stars, followed by axisymmetric NR simulations to study the post-merger evolution of the remnant system and ejecta formation~\citep{Fujibayashi20c, Fujibayashi23}. The subsequent evolution of the ejecta is followed by performing further hydrodynamics simulations employing the ejecta information obtained in these NR simulations, following all the way up to the time at which the homologous expansion of the matter profile is realized ($\sim 0.1\,{\rm days}$). 

The light curves are computed using a wavelength-dependent radiative transfer simulation code~\citep{Tanaka_2013,tanaka17,Tanaka:2017lxb, tanaka20,kawaguchi18,kawaguchi21}. In this code, the photon transfer is simulated by a Monte Carlo method for the ejecta profiles in the homologous expansion phase obtained by the hydrodynamics simulations, in combination with the elemental abundance and radioactive heating profiles obtained by the nucleosynthesis calculations. For the photon-matter interaction, the line list constructed by \citet{Domoto:2022cqp} are employed for the bound-bound transitions under the assumption of the local thermodynamic equilibrium (LTE). We refer to \citet{kawaguchi21, kawaguchi22, kawaguchi23} for the detailed setup of the radiative transfer simulations.

Among the models studied in~\cite{kawaguchi21, kawaguchi22, kawaguchi23}, we employ DD2-135, DD2-125, SFHo-135-135, and SFHo-120-150 as representative cases in this work. These models adopt different equations of state (EoS) and masses of merging neutron stars. DD2-135 and DD2-125 represent the cases in which the remnant massive neutron star survives for a long time scale ($\geq 1\,{\rm s}$), while SFHo-135-135 and SFHo-120-150 represent the cases in which the remnant massive neutron star collapses to a black hole within a short time scale ($\sim 10\,{\rm ms}$) after the onset of the merger. The key properties of the dynamical and post-merger ejecta for these four models are summarized in Table \ref{tab:model}. We note that there is no strict way to distinguish the dynamical and post-merger ejecta, as these components overlap to some extent. Therefore, we conventionally define the dynamical ejecta as the component of matter that is gravitationally unbound around the beginning of the axisymmetric NR simulations, and the post-merger component as the rest (see the caption of Table \ref{tab:model} for the details).

Figure~\ref{fig:sim-lc} presents the rest-mass density profile of the ejecta in the meridional plane for DD2-135 obtained by the hydrodynamics simulation at approximately 0.1 days, along with the $grizJHK$-band light curves, observed from both polar and equatorial viewing angles. In this model, the post-merger ejecta with a mildly prolate shape dominate the total ejecta mass, surrounded by the dynamical ejecta with an approximately spherical shape. While the $Y_e$ value of the post-merger ejecta is typically $\gtrsim 0.3$ with a weak spatial dependence, the dynamical ejecta have a significant $Y_e$ dependence on the latitudinal angle, $\theta$, with the values being lower and higher than 0.3 for $\theta\lesssim 45^\circ$ and $\gtrsim 45^\circ$, respectively. DD2-125 also has a similar property (see ~\citealt{kawaguchi21} and~\citealt{kawaguchi22} for more details). For SHFo-135-135 and SFHo-120-150, the morphology and $Y_e$ profiles of the dynamical and post-merger ejecta are similar to those in model DD2-135, but the total ejecta mass is smaller and the dynamical and post-merger ejecta have comparable masses (see \citealt{kawaguchi23}).

DD2-135 and DD2-125 show polar light curves similar to the observational data of GW170817/AT2017gfo, except for a more rapid decline in the optical wavelength after the peak in the theoretical light curves. On the other hand, SFHo-135-135 and SFHo-120-150 show the light curves much fainter than the observational data of GW170817/AT2017gfo due to the small ejecta mass (see also~\citealt{kawaguchi23} for the actual light curves). The light curves observed from the equatorial viewing angle are fainter than those from the polar viewing angle due to the presence of lanthanides \citep[e.g.,][]{Kasen:2014toa}, and such a viewing angle dependence is more pronounced in the optical wavelength.

\subsection{Parameter estimation}
\label{subsec:fitting}

\subsubsection{Preparation of the dataset}
We prepare the mock light curve data from the radiative transfer simulations with the following procedures.
To perform parameter estimation with an analytic model in the same way as done for the actual observational data, 
we first thin out the simulated light curve data. 
To create a dataset similar to actual observations,
we set the thinning interval to $0.5$ days. 
This choice of the constant interval may produce more data at later phase as compared with the actual observations.
However, we tested different intervals and confirmed that our conclusions are not affected by this choice: 
even when we thin the light curve more sparsely,
the results of the parameter estimation remain essentially unchanged. 
Additionally, we do not use the data in the initial phase ($< 0.5$ days) where the radiative transfer simulations are not reliable due to lack of opacity data.
Then, we impose a brightness limit to the simulated light curve. 
Since the faint parts of the light curves are often unobservable in actual observations, we limit the data with an absolute magnitude brighter than $ -12$ mag (see Section~\ref{sec:missing_data} for the impact of the missing observational data).

\subsubsection{Parameter estimation with analytic model}

\begin{table}[t]
  \begin{center}
  \caption{Prior distributions in the parameter estimation. For each parameter, we employ an uniform prior distribution between the minimum and maximum values given in the table.}
  \label{tab:prior}
  \hspace{-1cm}
    \begin{tabular}{ccc}
      \hline \hline
      Parameter & prior distribution & Units\\ \hline
      $M^{\rm{blue}}, M^{\rm{red}}$      & (0.001, 0.1) &  $\Msun$ \\
      $v^{\rm{blue}}, v^{\rm{red}}$      & (0.03, 0.7) &  $c$ \\
      $T_{\rm{c}}^{\rm{blue}}, T_{\rm{c}}^{\rm{red}}$  & (100, 4000) & $\rm{K}$ \\
      $\kappa^{\rm{red}}$  & (1.0, 30.0) &  $\rm{cm^2 g^{-1}}$\\
      $\sigma$                       & (0.1, 1.0) & $\rm{mag}$ \\
      \hline
    \end{tabular}
  \end{center}
\end{table}

We perform parameter estimation for the simulated light curve data with an analytic model using an ensemble-based Markov Chain Monte Carlo (MCMC) method. 
Our analytic model (see Appendix \ref{sec:model}) is broadly the same as that used by \cite{villar17},
 which is based on the prescription by \cite{metzger17} and implemented 
in \texttt{MOSFiT} \citep{nicholl17,villar17a, guillochon17}.

By following the common assumption, we employ two (blue and red) emission components.
An important assumption is that the blue and red emission components are modeled independently: the total flux is expressed as a simple sum of two components in each wavelength and no radiative reprocessing in between the components is considered.
We made the opacity of the blue component fixed at $\kappa^{\rm{blue}} = 0.5\ \rm{cm^2 g^{-1}}$ while we allow the opacity of the red component ($\kappa^{\rm{red}}$) to vary to align with the assumptions in \cite{villar17}. This model includes eight free parameters in total: two ejecta masses ($M^{\rm{blue}}, M^{\rm{red}}$), velocities ($v^{\rm{blue}}, v^{\rm{red}}$), and temperature floors ($T_c^{\rm{blue}}$, $T_c^{\rm{red}}$), one opacity ($\kappa^{\rm{red}}$), and one variance parameter ($\sigma$, see below). We apply flat prior distributions for these parameters, as shown in Table \ref{tab:prior}. Since these values are not necessarily tied to a physical interpretation (as we discuss below), we permit somewhat extreme values, in particular for velocities.

For the MCMC method, we utilize the Python package \texttt{emcee} \citep{Foreman-Mackey13}.
The form of the log-likelihood is:
\begin{equation}
  \begin{aligned}
    \ln\mathcal{L} = \frac{1}{2}\sum_{i = 1}^{n}\left[\frac{(O_{i,\rm{sim}} -O_{i,\rm{model}})^2}{\sigma_i^2+\sigma^2}
    - \ln(2\pi\sigma_i^2)\right]\\
    - \frac{n}{2}\ln(2\pi\sigma^2)
  \end{aligned} 
\end{equation}
where $O_{i,\rm{sim}}$ and $O_{i,\rm{model}}$ represent the $i^{\rm{th}}$ of $n$ simulated magnitudes and analytic model magnitudes, respectively. In the case of actual observations, $\sigma_i$ represents the error for each photometric point. For our simulation data, we fix the value at $\sigma_i =0.2\ \mathrm{mag}$. Additionally, we introduce the variance parameter $\sigma$ which represents the additional uncertainty in the analytic model and/or simulation data. This $\sigma$ is one of the parameters estimated through MCMC.

To confirm the validity of our methods, we fit the the observational data of GW170817/AT2017gfo \citep{villar17}. The model realizations with the highest likelihood (“best-fit”) show a good agreement with the observational data. 
Inferred parameters both for blue and red components also show broad agreement with those by \cite{villar17} (see Appendix \ref{sec:obsfit} for more details).


\section{Results}
\label{sec:result}

\begin{table*}[ht]
  \begin{center}
  \caption{Inferred parameters for different models}
  \label{tab:result}
  \makebox[1 \textwidth][c]{
    \scalebox{0.9}{
    \hspace{-2.5cm}
    \begin{tabular}{llccccccccc}
      \hline \hline
      Model & viewing angle & $M^{\rm{blue}}$ & $v^{\rm{blue}}$ & $\kappa^{\rm{blue}}$ & $T_c^{\rm{blue}}$ & $M^{\rm{red}}$ & $v^{\rm{red}}$ & $\kappa^{\rm{red}}$ & $T_c^{\rm{red}}$ & $\sigma$ \\ \hline
      DD2-135 & polar (fiducial) & 
      $0.010_{-0.001}^{+0.001}$ & $0.43_{-0.02}^{+0.02}$ & (0.5) & $3100_{-505}^{+246}$ & $0.028_{-0.001}^{+0.001}$ & $0.26_{-0.01}^{+0.02}$ & $3.4_{-0.2}^{+0.1}$ & $1834_{-44}^{+46}$ & $0.102_{-0.001}^{+0.003}$ \\
      DD2-135\_e & equatorial & 
      $0.004_{-0.000}^{+0.000}$ & $0.46_{-0.03}^{+0.03}$ & (0.5) & $3546_{-184}^{+204}$ & $0.022_{-0.001}^{+0.001}$ & $0.25_{-0.01}^{+0.02}$ & $4.6_{-0.2}^{+0.2}$ & $1851_{-50}^{+51}$ & $0.102_{-0.001}^{+0.003}$ \\
       \hline
      DD2-135\_pm03 & polar & 
      $0.005_{-0.000}^{+0.000}$ & $0.44_{-0.02}^{+0.02}$ & (0.5) & $3408_{-522}^{+247}$ & $0.011_{-0.001}^{+0.001}$ & $0.25_{-0.02}^{+0.02}$ & $4.3_{-0.2}^{+0.3}$ & $1919_{-62}^{+65}$ & $0.102_{-0.002}^{+0.004}$ \\
      DD2-135\_pm01 & polar & 
      $0.003_{-0.000}^{+0.000}$ & $0.44_{-0.03}^{+0.03}$ & (0.5) & $3317_{-310}^{+348}$ & $0.004_{-0.000}^{+0.000}$ & $0.23_{-0.02}^{+0.03}$ & $4.9_{-0.5}^{+0.5}$ & $1980_{-124}^{+108}$ & $0.104_{-0.003}^{+0.010}$ \\
      \hline
      DD2-125 & polar & 
      $0.009_{-0.001}^{+0.001}$ & $0.36_{-0.03}^{+0.03}$ & (0.5) & $2725_{-625}^{+433}$ & $0.051_{-0.003}^{+0.003}$ & $0.22_{-0.02}^{+0.02}$ & $2.5_{-0.2}^{+0.2}$ & $1660_{-55}^{+62}$ & $0.329_{-0.031}^{+0.036}$\\
      SFHo135-135 & polar & 
      $0.003_{-0.000}^{+0.000}$ & $0.61_{-0.04}^{+0.04}$ & (0.5) & $3426_{-536}^{+455}$ & $0.011_{-0.001}^{+0.001}$ & $0.38_{-0.03}^{+0.03}$ & $5.7_{-0.3}^{+0.3}$ & $1285_{-71}^{+71}$ & $0.103_{-0.002}^{+0.006}$ \\
      SFHo120-150& polar & 
      $0.001_{-0.000}^{+0.000}$ & $0.57_{-0.03}^{+0.04}$ & (0.5) & $2361_{-1043}^{+679}$ & $0.016_{-0.001}^{+0.001}$ & $0.34_{-0.02}^{+0.02}$ & $5.6_{-0.2}^{+0.2}$ & $1388_{-52}^{+60}$ & $0.102_{-0.002}^{+0.004}$ \\ 
      \hline
    \end{tabular}
    }
  }
  \end{center}
\end{table*}

\begin{figure*}  
\begin{center}
  \begin{tabular}{cc}
    \includegraphics[width=7.5cm]{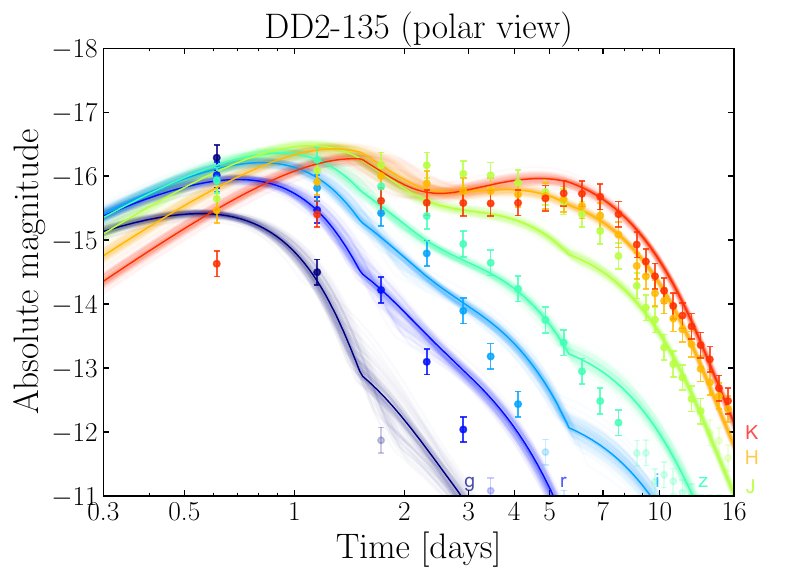}&
    \includegraphics[width=7.5cm]{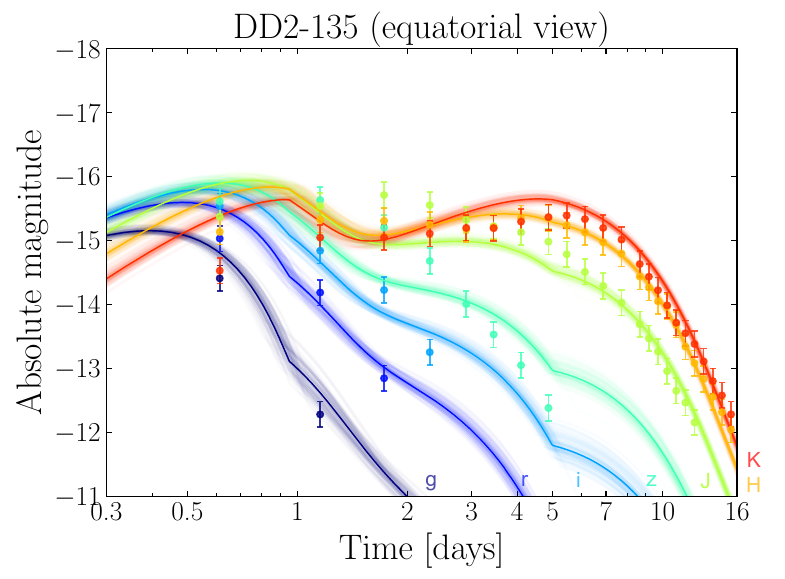}
  \end{tabular}
\caption{(Left) Mock observational data prepared by radiative transfer simulation: DD2-135 viewed  from a polar angle ($0^\circ<\theta<20^\circ$, points). Thick points are the data used for parameter estimation. Solid lines represent the realizations of the highest likelihood (best-fit) for each filter, while thin lines show the projections of results from 100 randomly chosen chains. (Right) The same as the left panel but for the light curves viewed from the equatorial angle ($86^\circ<\theta<90^\circ$).
}
\label{fig:fiducial-lc}
\end{center}
\end{figure*}

\begin{figure*}
\begin{center}
\includegraphics[width=12cm]{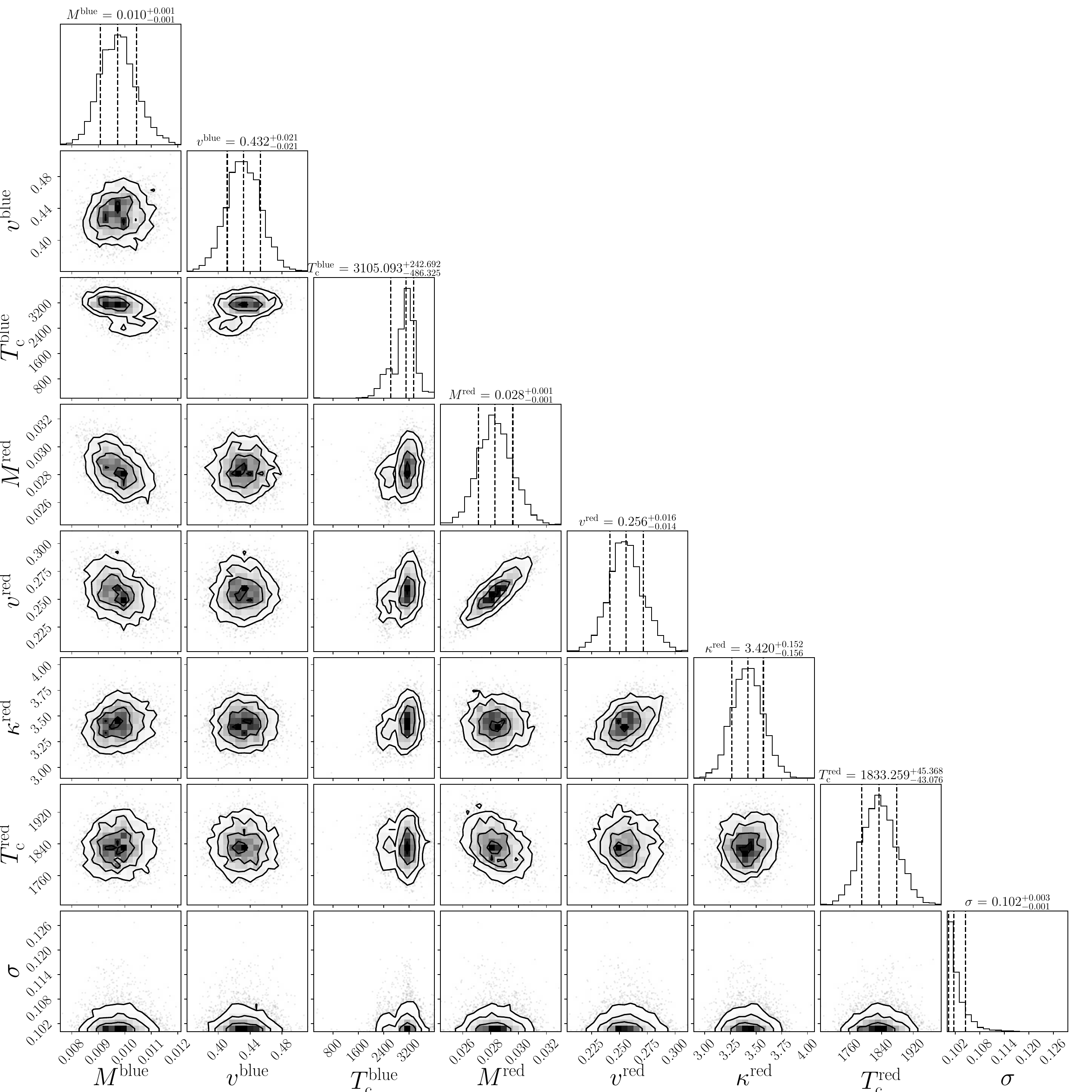}
\caption{Corner plot showing the posterior distributions of parameters obtained by using our method for the fiducial case (DD2-135, polar view).
}
\label{fig:fiducial-corner}
\end{center}
\end{figure*}

\begin{figure*}  
  \begin{center}
    \begin{tabular}{cc}
      \includegraphics[width=8cm]{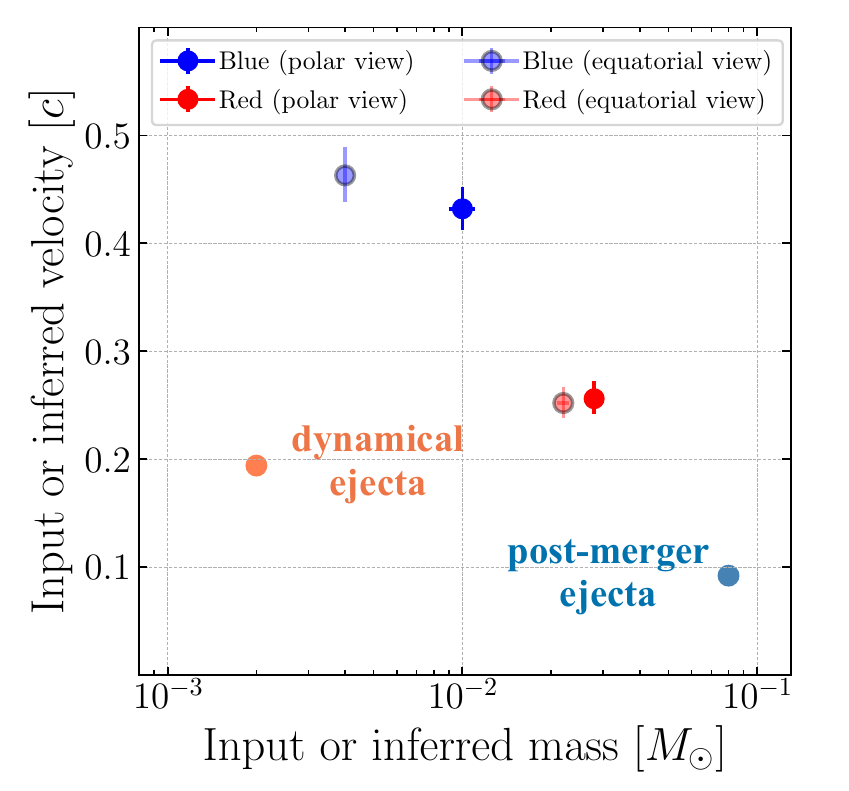} &
      \hspace{-1.0cm}
      \includegraphics[width=10.5cm]{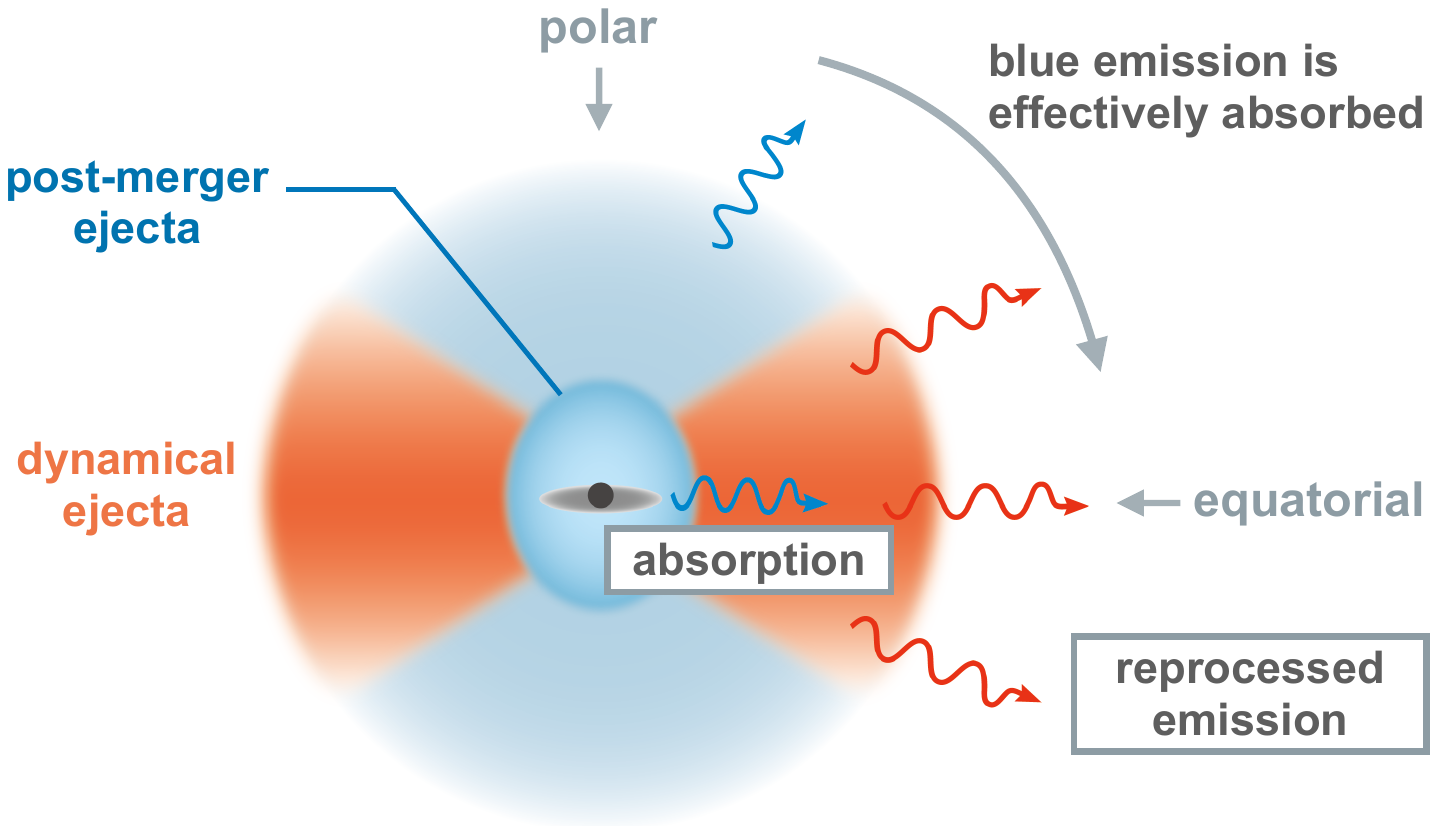}
    \end{tabular}
  \caption{(Left) Relationship between the mass and velocity for input values of DD2-135 ($M_{\rm{ej}}^{\rm{dyn}}, M_{\rm{ej}}^{\rm{dyn}}$), ($M_{\rm{ej}}^{\rm{pm}}, M_{\rm{ej}}^{\rm{pm}}$) and parameters inferred from the analytic modeling ($M^{\rm{blue}}, v^{\rm{blue}}), (M^{\rm{red}}, v^{\rm{red}}$). Thick circles represent the estimated parameters for the light curves viewed from the polar angle, while thin circles represent those for the light curves viewed from the equatorial angle. 
  (Right) Schematic picture of the photon transfer in kilonova ejecta. In the polar direction, blue emission from the post-merger ejecta (lower $Y_e$, lower velocity) can escape efficiently. In the equatorial direction, on the other hand, the emission from the post-merger ejecta is absorbed by the dynamical ejecta (higher $Y_e$, higher velocity) and reprocessed to red emission.
}
  \label{fig:result-Mv}
  \end{center}
\end{figure*}

\begin{figure*}  
  \begin{center}
    \begin{tabular}{cc}
      \includegraphics[width=8cm]{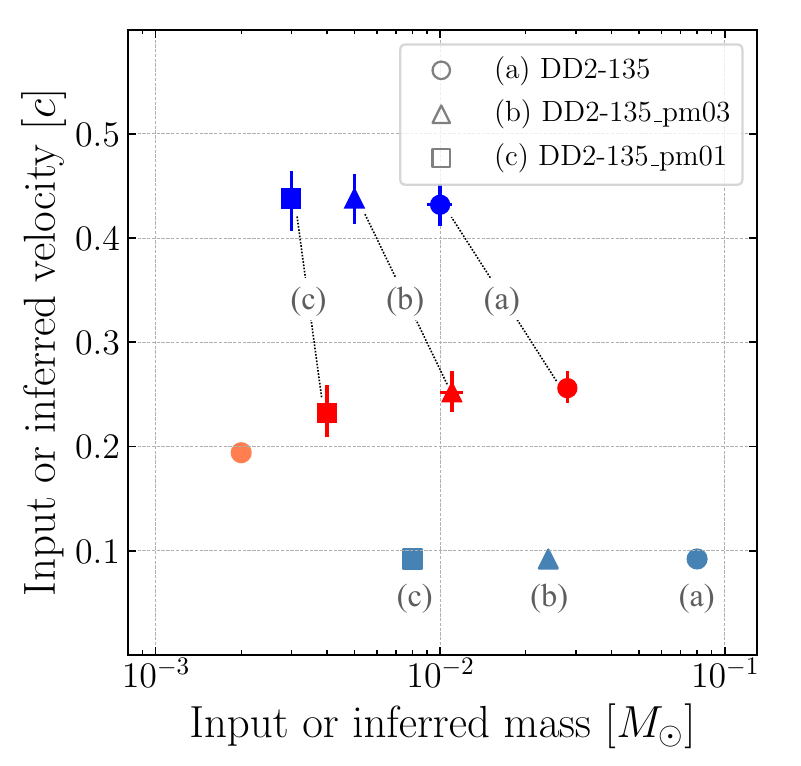}
    \end{tabular}
  \caption{Relationship between the mass and velocity for three simulations: (a) the fiducial model (DD2-135), and two additional simulation models (b) DD2-135\_pm03 and (c) DD2-135\_pm01 with 30\% and 10\% of the post-merger ejecta mass as compared with the fiducial model. Although these models adopt the same dynamical ejecta, the mass of both blue and red components are affected by changing the mass of the post merger ejecta.
}
  \label{fig:Mv-pm}
  \end{center}
\end{figure*}

\begin{figure*}  
  \begin{center}
    \begin{tabular}{cc}
      \includegraphics[width=18cm]{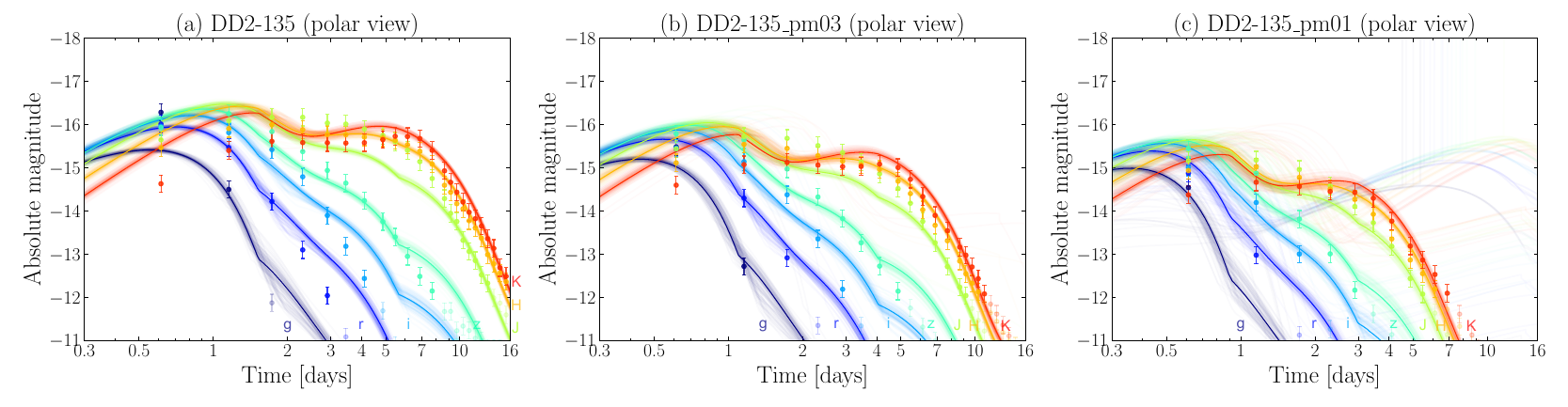}
    \end{tabular}
  \caption{Best-fit multiband light curves of three models in Figure \ref{fig:Mv-pm}. Although only the mass of the post-merger ejecta is modified across the models, the brightness of the red emission (i.e., NIR emission $t > 3$ days) is also affected.
}
  \label{fig:fiducial_lcs}
  \end{center}
\end{figure*}

We perform parameter estimation from the simulation data with the analytic light curve model. As the fiducial case, we use the light curve of DD2-135 observed from the polar angle, as it closely resembles the light curves of GW170817/AT2017gfo (Figure \ref{fig:sim-lc}). First, we present the results for the fiducial case and compare the inferred parameters with the actual ejecta properties from the NR simulations
(Section \ref{subsec:fiducial}).
Also, to understand the physical meaning of the inferred parameters, we perform two additional radiative transfer simulation by varying post-merger ejecta masses (Section \ref{subsec:pmmass}).
In addition, we study the dependence of the parameter estimation on viewing angles (Section \ref{subsec:viewingangle}) as well as merger models (Section \ref{subsec:models}). A summary of the parameter estimation is presented in Table \ref{tab:result}.

\subsection{Fiducial case}
\label{subsec:fiducial}
The best-fit light curves for our fiducial case are shown in the left panel of Figure \ref{fig:fiducial-lc}.
The corresponding corner plot is shown in Figure \ref{fig:fiducial-corner}. 
Overall, the analytic model provides a good agreement with the simulation data. Similar to the two-component modeling of the observational data of GW170817/AT2017gfo, 
the early phase of the light curve in the simulation data is reproduced by the blue component, while the later phase is reproduced by the red component.

The left panel in Figure \ref{fig:result-Mv} shows the relationship between the masses and velocities.
The figure compares parameters estimated from analytic modeling (($M^{\rm{blue}}, \ v^{\rm{blue}}$), ($M^{\rm{red}}, \ v^{\rm{red}}$)) and input parameters in the radiative transfer simulation (($M_{\rm{ej}}^{\rm{dyn}}, \ v_{\rm{ej}}^{\rm{dyn}}$), ($M_{\rm{ej}}^{\rm{pm}}$, $\ v_{\rm{ej}}^{\rm{pm}}$)). 
Here input parameters mean those obtained by the NR simulations, 
which are adopted as the initial conditions of the radiative transfer simulations.
If the dynamical and post-merger ejecta correspond to the red and blue emission components, respectively, 
the input and estimated mass and velocity should agree with each other.
However, as shown in Figure \ref{fig:result-Mv}, the estimated parameters largely deviate from the input parameters: 
the input values are 
$M_\mathrm{ej}^\mathrm{dyn}=0.0015\ \Msun$,
$v_\mathrm{ej}^\mathrm{dyn}=0.19\ c$, 
$M_\mathrm{ej}^\mathrm{pm}=0.080\ \Msun$, and
$v_\mathrm{ej}^\mathrm{pm}=0.092\ c$, while
the best-fit parameters are $M^\mathrm{blue}=0.010_{-0.001}^{+0.001}\ \Msun$, $v^\mathrm{blue}=0.43_{-0.02}^{+0.02}\ c$, $M^\mathrm{red}=0.028_{-0.001}^{+0.001}\ \Msun$ and $v^\mathrm{red}=0.26_{-0.01}^{+0.02}\ c$ (see Table \ref{tab:result}).

Notably, under the naive assumption that dynamical and post-merger ejecta correspond to the red and blue components, respectively, the hierarchy in the estimated masses and velocities is opposite from those in the input values.
The NR simulations generally predict that dynamical ejecta are less massive and have a higher velocity, i.e., $M_{\rm{ej}}^{\rm{dyn}} < M_{\rm{ej}}^{\rm{pm}}$ and $v_{\rm{ej}}^{\rm{dyn}} > v_{\rm{ej}}^{\rm{pm}}$. 
On the other hand, the results of analytic modeling show that the red component is more massive and has a lower velocity, 
i.e., $M^{\rm{red}}> M^{\rm{blue}}$ and $v^{\rm{red}} < v^{\rm{blue}}$. 
In fact, the trend in the estimated parameters for the blue and red components are consistent with the findings from two-component modeling of the observational data of GW170817/AT2017gfo \citep[e.g.,][]{villar17}. 
Our results clearly indicate that the parameters estimated by the analytic light curve modeling do not represent the actual configuration of the kilonova ejecta. 

\subsection{Physical picture of the emission}
\label{subsec:pmmass}

To understand the physical meaning of the estimated parameters, we peform two additional radiative transfer simulations with different post-merger ejecta masses: DD2-135\_pm03 and DD2-135\_pm01 with 30\% and 10\% of the mass in the fiducial model, respectively. 
In Figure \ref{fig:Mv-pm}, we show the results of parameter estimation for the fiducial case and these two additional cases. In all the cases, we use the light curves observed from polar angle. We find that, by decreasing the mass of the post-merger ejecta, the masses of both blue and red components are reduced. 
As shown in the light curves of these cases (Figure \ref{fig:fiducial_lcs}), the reduction in the post-merger ejecta mass affects both blue-early and red-late emission,
which clearly indicate that the post-merger ejecta contribute to both blue and red emission components. 

The physical picture of the emission from two ejecta components is summarized in the right panel of Figure \ref{fig:result-Mv}.
The dynamical ejecta is ejected with a faster velocity than the post-merger ejecta, 
and they are primarily located in the equatorial plane. 
The post-merger ejecta are ejected later with a slower velocity.
Due to the higher $Y_e$ (lanthanide-poor), the post-merger ejecta mainly emits bluer emission. 
In the equatorial direction, the blue emission is effectively absorbed and reprocessed to red emission by the dynamical ejecta, which tend to have a lower $Y_e$ (lanthanide-rich).

Based on this physical picture, we can approximately understand the 
relation between the estimated mass from the physical mass of each ejecta
by introducing a correction factor.
For example, we introduce a surface covering factor of the lanthanide-rich dynamical ejecta $f_{\Omega}$. (Note that this parameter is not used in our analytic light curve modeling.)

In our fiducial model (DD2-135), the surface covering factor of the lanthanide-rich dynamical ejecta is $f_{\Omega} \sim 0.6-0.7$.
Thus, approximately $1-f_{\Omega}$ of the emission from the post-merger ejecta is observed as the blue emission for the polar direction while the remaining is reprocessed to the red emission.
In this simple picture, by using $M^{\rm pm}_{\rm ej} = 0.080 \ \Msun$ and $M^{\rm dyn}_{\rm ej} = 0.0015 \ \Msun$ in DD2-135, the estimated mass of the blue component would be 
$M^{\rm blue} \sim (1-f_{\Omega}) M^{\rm pm}_{\rm ej}
\sim 0.03 \ \Msun$,
while that of the red component would be
$M^{\rm red} \sim f_{\Omega} M^{\rm pm}_{\rm ej} + M^{\rm dyn}_{\rm ej} 
\sim 0.05 \ \Msun$.
These naive estimates neglect the difference in the surface area of blue and red emission components.
Nevertheless, this simple picture naturally explains the behaviors in the estimated masses, i.e., $M^{\rm red} > M^{\rm blue}$.
Note that the estimated total mass is different from our results of parameter estimation due to the difference in specific heating rate (see Section \ref{sec:discussion}).

Since the post-merger ejecta has a mass comparable to or even higher than that of the dynamical ejecta in variety of mergers with different neutron star masses, the post-merger ejecta has a significant contribution for powering the red emission.
Thus, we cannot naively connect the inferred mass of the red component with the mass of the dynamical ejecta.
Furthermore, it is challenging to estimate the mass of the dynamical ejecta when its contribution as a heating source is minor,
i.e., $f_{\Omega} M^{\rm pm}_{\rm ej} > M^{\rm dyn}_{\rm ej}$.

The estimate of the mass also largely affects the estimate of the velocity as the velocity is only the parameter in the analytic model to control the timescale of the emission for a given mass and opacity.
The estimated velocities are generally higher than those of the input values (Figures \ref{fig:result-Mv} and \ref{fig:Mv-pm}).
The high velocity of the blue component is required to reproduce the early light curve peak in the optical wavelengths as well as their subsequent rapid fading with the small estimated mass and temporally constant gray opacity (see~\citealt{kawaguchi21} for the effect of the time/wavelength dependent opacity in the late phase optical light curves).
Also, the high velocity of the red component is required to reproduce the peak time of the red emission by compensating with the large estimated mass.
As a result, similar to the situations in the ejecta mass, the estimated velocities for each component do not correspond to the actual velocities of the ejecta.

\subsection{Dependence on viewing angle}
\label{subsec:viewingangle}

The kilonova light curve can vary also due to the viewing angle. To study the dependence of the inferred parameters on the viewing angle, 
we also perform the parameter estimation for the light curve of the model DD2-135 observed from the equatorial angle. 
We show the best fit light curves in the right panel of Figure \ref{fig:fiducial-lc} and inferred parameters in the left panel of Figure \ref{fig:result-Mv}.
The best-fit parameters are $M^\mathrm{blue}=0.005_{-0.001}^{+0.001}\ \Msun$, $v^\mathrm{blue}=0.46_{-0.03}^{+0.03}\ c$, $M^\mathrm{red}=0.022_{-0.001}^{+0.001}\ \Msun$ ,and $v^\mathrm{red}=0.25_{-0.01}^{+0.02}\ c$. 

As compared with the fiducial case (observed from the polar angle), the estimated mass is decreased for the equatorial viewing angle:
the decrease is $\sim 60\%$ for the blue component and $\sim 21\%$ for the red component. 
The decrease in the total mass results from the fact that more emission escapes toward polar direction due to the anisotropic distribution of the ejecta \citep{Kasen15, kawaguchi20, Darbha&Kasen20, Korobkin21, kawaguchi21}.
The greater reduction in the blue component compared to the red component can also be understood by the presence of anisotropic dynamical ejecta. 
For the equatorial direction, blue emission from the post-merger ejecta is effectively absorbed by the lanthanide-rich dynamical ejecta located in the equatorial plane.
Only a part of the blue emission from the polar dynamical ejecta and the prolate post-merger ejecta contribute to the equatorial observers.
In other words, the surface area of the blue emission toward the equatorial observer is significantly reduced. 
As a result, the blue emission is largely suppressed in the equatorial direction, which results in the lower mass of the blue component (see the right panel of Figure \ref{fig:result-Mv}).

\subsection{Dependence on merger models}
\label{subsec:models}

\begin{figure*}
  \begin{center}
    \begin{tabular}{cc}
      \includegraphics[width=7cm]{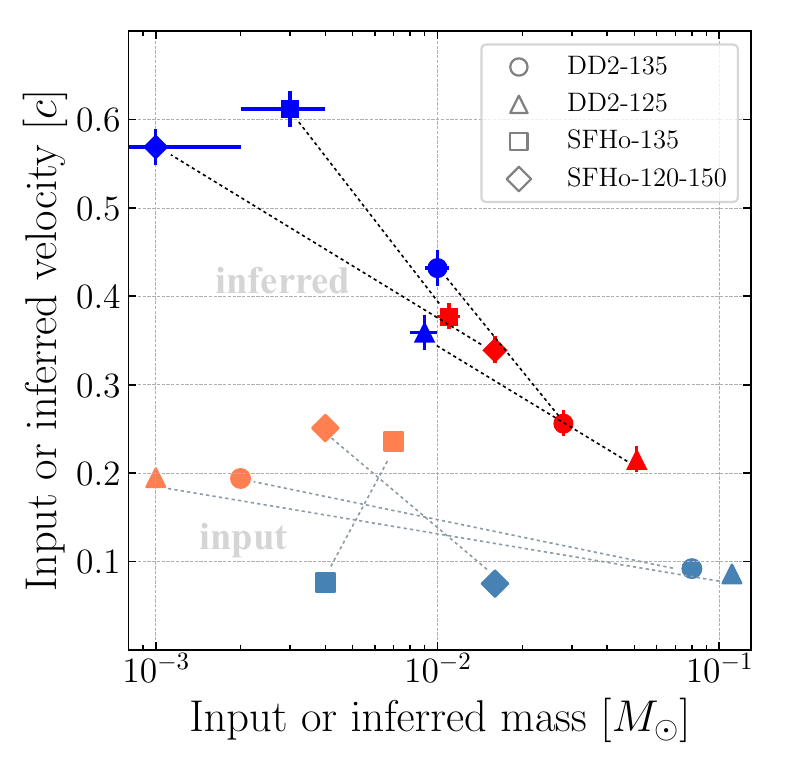} &
      \hspace{-0.5cm}
      \includegraphics[width=11.5cm]{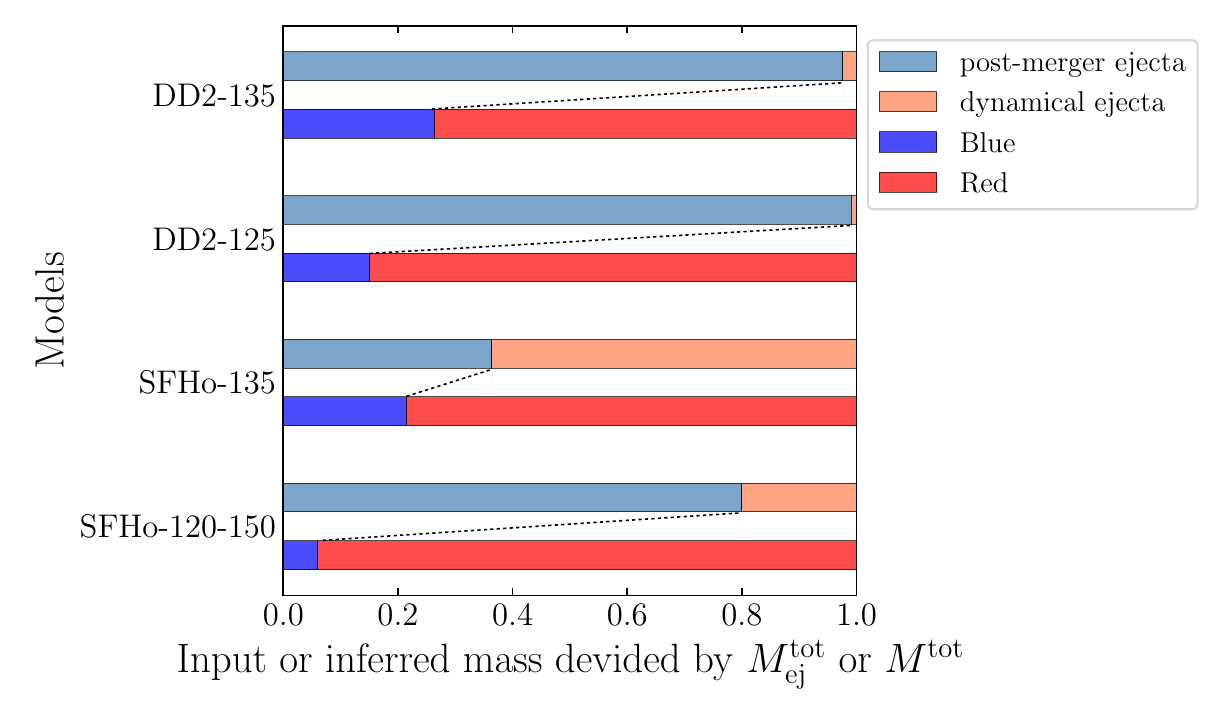}
    \end{tabular}
  \caption{(Left) Relationship between the mass and velocity for four different merger models and the inferred parameters from the light curve modeling. Each symbol represents a different model. (Right) The mass ratio of the post-merger/dynamical (input parameters) and blue/red components (estimated parameters) for each model. Each bar is normalized by the total mass, i.e.,   $M^{\mathrm{tot}}_{\rm{ej}}=M_{\rm{ej}}^{\rm{dyn}}+M_{\rm{ej}}^{\rm{pm}}$ for the input values (upper bars) and 
  $M^{\mathrm{tot}}=M^{\rm{blue}}+M^{\rm{red}}$ for the inferred parameters (bottom bars).
  }
  \label{fig:4model}
  \end{center}
\end{figure*}

In Figure \ref{fig:4model}, we compare the results of parameter estimation for four different merger models with different neutron star masses and EOSs. The left panel shows the relationship between mass and velocity. The circle points represent our fiducial model, while other symbols correspond to the other models listed in Table \ref{tab:model}. Overall, we find the same trend as in the fiducial model: for all the cases, the relationships of the parameters show $M^{\rm{red}} > M^{\rm{blue}}$ and $v^{\rm{red}} < v^{\rm{blue}}$.

The right panel of Figure \ref{fig:4model} focuses on the mass ratio for each model.
Each bar is normalized by the total ejecta mass $M^{\mathrm{tot}}_{\rm{ej}} \ (=M_{\rm{ej}}^{\rm{dyn}}+M_{\rm{ej}}^{\rm{pm}})$ or the sum of the inferred masses of the blue and red components $M^{\mathrm{tot}} \ (=M^{\rm{blue}}+M^{\rm{red}})$.
We find that, for the models where the post-merger ejecta dominates ($M_{\rm{ej}}^{\rm{pm}} > M_{\rm{ej}}^{\rm{dyn}} $), the estimated mass of the blue component is always smaller than that of the red component
($M^{\rm{blue}} < M^{\rm{red}}$). 
These behaviors can be understood in the same way as in the fiducial case: the emission from the post-merger ejecta contributes to both blue and red emission.

An exception is the model SFHo-135, where the dynamical ejecta is more massive than the post-merger ejecta ($M_{\rm{ej}}^{\rm{dyn}} > M_{\rm{ej}}^{\rm{pm}}$).  
In this case, the derived parameters follow $M^{\rm{red}} > M^{\rm{blue}}$ (as in the other cases).
However, the proportion of the total mass occupied by the red mass is greater than that occupied by the dynamical ejecta. This is also explained by the same effect seen in the other three models: not only the dynamical ejecta but also the post-merger ejecta contributes to the red emission through reprocessing.
We emphasize that the apparent discrepancy between the input values and inferred parameters exists in general, not just in a specific model.


\section{Discussion}
\label{sec:discussion}

\subsection{Interpretations of estimated parameters}

\begin{figure}
  \begin{center}
  \hspace{-1.2cm}
  \includegraphics[width=7cm]{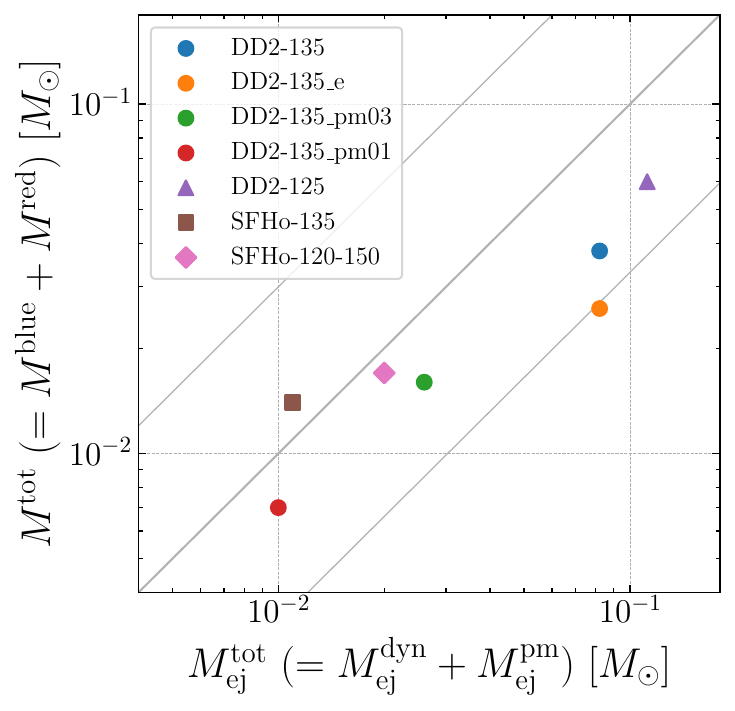}
  \caption{Relationship between the total ejecta mass in the simulations $M^{\mathrm{tot}}_{\rm{ej}}\ (=M_{\rm{ej}}^{\rm{dyn}}+M_{\rm{ej}}^{\rm{pm}})$ and the sum of the inferred masses of the blue and red components $M^{\mathrm{tot}}\ (=M^{\rm{blue}}+M^{\rm{red}})$. The lines represent $M^{\mathrm{tot}}=M^{\mathrm{tot}}_{\rm{ej}}$ and a factor of 3 difference.
  }
  \label{fig:emmass}
  \end{center}
\end{figure}

By using the light curves obtained from multi-dimensional radiative transfer simulations, we perform parameter estimation with a commonly-used analytic light curve model.
The estimated parameters for the fiducial model (DD2-135, viewed from the polar direction)
are similar to those obtained for the observational data of GW170817/AT2017gfo (Section \ref{subsec:fiducial}). 
We find that the parameters do not correspond to the input parameters of radiative transfer simulations. This indicates that the parameters estimated by analytic light curve modeling do not represent the actual configuration of the kilonova ejecta.
In other words, our results demonstrate that the “red component” and “blue component” in the analytic model are not directly connected to the physical components of the ejecta.

As shown in Figure \ref{fig:Mv-pm} and Figure \ref{fig:fiducial_lcs}, the post-merger ejecta contributes to both blue and red emissions. 
This is because the blue emission from the (typically more massive) post-merger ejecta is absorbed by the dynamical ejecta and reprocessed to the red emission (see the right panel of Figure \ref{fig:result-Mv}).
Also we demonstrate that the inferred parameters depend on the viewing angle:
the blue emission tends to be suppressed (resulting in a lower inferred $M^{\rm{blue}}$) when observed from the equatorial angle. This result is also a natural outcome of the physical picture outlined above.

In fact, it is known that a similar two component light curve model works very well for supernovae \citep[e.g.,][]{maeda03,valenti08}.
A striking difference between supernovae and kilonovae is their ejecta structure and opacities. For supernovae, two-component models are invoked to express the emission from the denser inner ejecta and from the surrounding outer ejecta. 
These two parts have a similar opacity as the compositions are not very different.
The diffusion timescale of the core is longer than that of the outer ejecta.
In this case, the emission from the core becomes important after the outer ejecta becomes optically thin.
Thus, a simple sum of the two fluxes is a fair assumption.
On the other hand, for the case of kilonovae, post-merger ejecta are surrounded by the dynamical ejecta.
Due to the large difference in the opacities, the diffusion timescale of the central post-merger ejecta can be shorter. Then, the emission from the post-merger ejecta is absorbed and reprocessed by the dynamical ejecta with a high opacity.
In addition, the lanthanide-rich dynamical ejecta is not spherical, and some of the emission from post-merger ejecta can directly leak toward polar region.
As a result, a simple sum of the flux from two components does not represent the actual emission for kilonovae.

Then, a question is which physical quantities can be robustly estimated from the analytic modeling of kilonova light curve.
Figure \ref{fig:emmass} shows the relationship between the total ejecta mass of the simulations $M^{\mathrm{tot}}_{\rm{ej}}\ (=M_{\rm{ej}}^{\rm{dyn}}+M_{\rm{ej}}^{\rm{pm}})$ and the sum of the inferred masses of the blue and red components $M^{\mathrm{tot}}\ (=M^{\rm{blue}}+M^{\rm{red}})$. 
It is shown that the sum of the estimated masses recover the actual ejected mass in the simulations within a factor of about $3$. 
This is natural because the total luminosity of kilonova is controlled by the total ejecta mass.
According to current NR simulations for a variety of mergers with different neutron star masses, the post-merger ejecta mass often dominates the total ejecta mass or is comparable to the dynamical ejecta mass  \citep{Fujibayashi20c, Fujibayashi23, Just23} . Thus, estimating the total mass provides a rough indication of the post-merger ejecta.

Note that for DD2-135 and DD2-125 the total inferred mass is less than the total input mass.  This discrepancy arises from differences in the heating rate. 
Our analytic model assumes the same commonly-adopted specific heating rate regardless of the ejecta properties.
In these two merger models, however, the post-merger ejecta with a high $Y_e$ dominates the total ejecta mass, leading to a lower specific heating rate \citep{Wanajo14, Wu16, Lippuner17, kawaguchi21}.  
As a result, the inferred masses from the analytic modeling tend to be smaller as compared with the actual ejecta mass.

\subsection{Impacts of missing observational data}\label{sec:missing_data}

\begin{figure*}  
  \begin{center}
    \begin{tabular}{cc}
      \includegraphics[width=17cm]{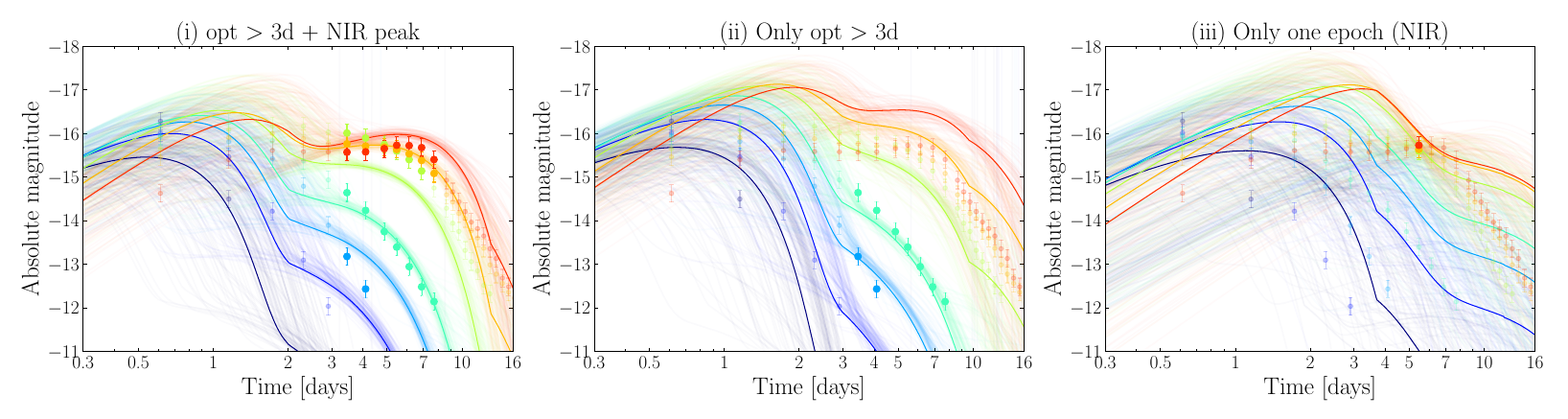}
    \end{tabular}
  \caption{Best-fit multiband light curves for three datasets with different observational conditions: (i) Observed in optical bands ($i, z$) starting at 3 days after the merger, and observed in the NIR bands ($J, H, K$) only for points brighter than $-15$ mag. (ii) Observed only in the red optical bands ($i, z$) starting at 3 days after the merger. (iii) Only one epoch (one point for each NIR band ($J, H, K$)) are taken at 5 days after the merger. Thick points are the data used for parameter estimation in each case. 
}
  \label{fig:deficit_lcs}
  \end{center}
\end{figure*}

\begin{figure}[ht]
  \begin{center}
    \begin{tabular}{cc}
    \hspace{-1.5cm}
      \includegraphics[width=6cm]{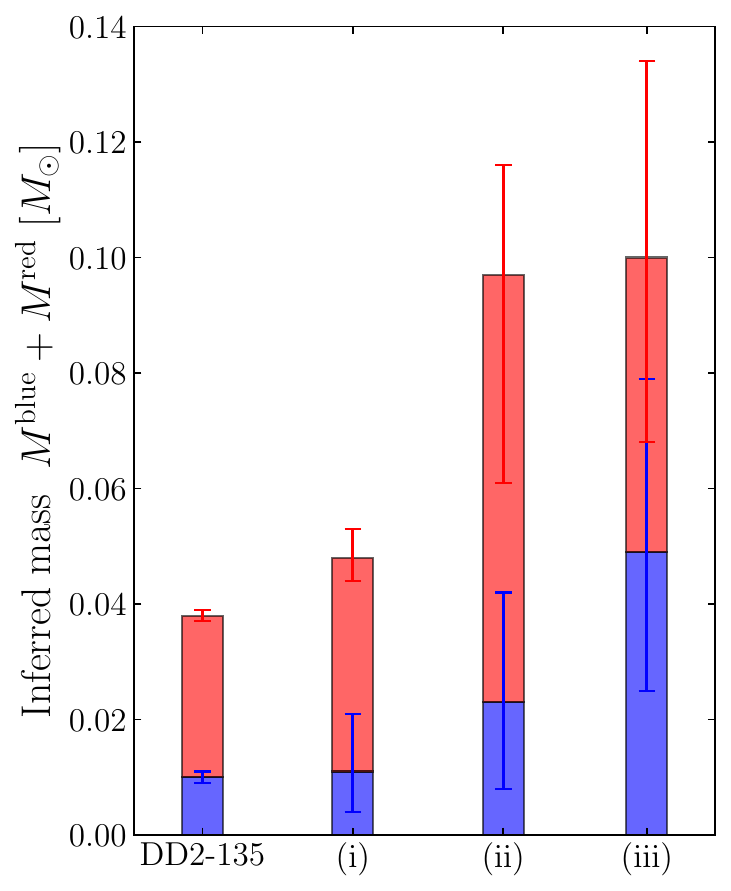}
    \end{tabular}
    \caption{Estimated masses for three datasets with different observational conditions in Figure \ref{fig:deficit_lcs}, compared to the results with complete data.
  }
\label{fig:deficit}
\end{center}
\end{figure}

As discussed in the previous section, our results suggest that the estimate of the total ejecta mass is relatively robust.
However, in actual observations, it can be challenging to obtain a complete light curve. When some data are missing, parameter estimation should be performed using only the limited light curves (e.g., \citealt{O'Connor21, Rastinejad22, Rastinejad24, Yang24}). To demonstrate that incompleteness of the data has in fact a large effect on the outcomes of parameter estimation, we perform parameter estimation by removing a part of the simulated light curve for model DD2-135. We find that the estimated mass tends to be larger when some data are missing, due to difficulties in defining the peak magnitude of the light curves.

We investigate several cases of observational conditions. 
Figure \ref{fig:deficit_lcs} shows three cases representing a possible situations where data might be incomplete:
(i) Observed in the red optical bands ($i, z$) starting at 3 days after the merger, and observed in the NIR bands ($J, H, K$) only for points brighter than $-15$ mag.
(ii) Observed only in the red optical bands ($i, z$) starting at 3 days after the merger.
(iii) Only one epoch (one point for each NIR band ($J, H, K$)) are taken around 5 days after the merger.
Cases (i) and (ii) assume scenarios where there is a delay in initiating follow-up observations triggered by gravitational wave detection. In case (i), follow-up in NIR bands is possible, whereas it is not in case (ii).
Case (iii) assumes a situation of kilonova observations following the detection of GRBs. In such cases, the early optical light curve are often dominated by the afterglow of GRBs. In some instances, observations may be limited to the NIR wavelength range only, as seen in GRB 130603B \citep{berger13, tanvir13} and GRB 200522A \citep{Fong21, O'Connor21}.

Figure \ref{fig:deficit} shows estimated masses from analytic modeling of these three conditions, as compared to the case with complete data. 
In case (i), with the optical data points after the peak and NIR data points near the peak, the results of the parameter estimation remain almost unchanged.
In contrast, in case (ii) with only optical data, 
the estimated mass is about twice as large as in the case of complete data. 
Since the latter part of the kilonova light curves is dominated in the NIR wavelengths, lack of NIR data leaves the total luminosity at later phase unconstrained (see case (ii) in Figure \ref{fig:deficit_lcs}). 
This results in the uncertainty in the estimate of the total mass.
Similarly, in case (iii), with only one epoch in NIR bands, the inferred mass is also about twice as large as in the case with complete data.
This is because the exact peak of the NIR light curve (the total luminosity at the later phase) is not constrained (see case (iii) in Figure \ref{fig:deficit_lcs}).
Since we cannot study all possible cases of missing data, these results should be considered just as examples. 
Nevertheless, to achieve better recovery of the ejecta parameters, 
it is important to perform observations to capture the emission in multiple NIR bands near the peak.

\section{Summary}
We study the physical meaning of the parameters estimated from analytic modeling of kilonova light curves. Using the results of radiative transfer based on NR simulations \citep{kawaguchi21, kawaguchi22, kawaguchi23}, we perform parameter estimation by analytically modeling the simulated light curves. Our results show that the estimated parameters for the blue and red components do not represent the input properties of the dynamical and post-merger ejecta adopted in the simulations, respectively. This discrepancy is commonly seen across different merger models with different neutron star masses and EOSs.

The main reason of this behavior is the reprocessing of photons from post-merger ejecta. 
In the mergers of a variety of neutron star masses, 
the post-merger ejecta tend to have a mass comparable to or higher than the dynamical ejecta, and thus, the post-merger ejecta serves as the significant heating source.
A part of this emission from the post-merger ejecta is absorbed by the dynamical ejecta, which is mainly distributed in the equatorial plane. 
As a result, the photon is reprocessed to red emission, which is interpreted as the red component in the analytic light curve model.
Our results caution against discussing separately the origins of ``red'' and ``blue'' components inferred by the analytic models that simply superposes the fluxes of two components.

We also study viewing angle dependence of the estimated parameters. 
When a kilonova is observed from the equatorial direction, the inferred mass of the blue component is largely suppressed.
This is also understood as a consequence of photon reprocessing by the dynamical ejecta.

Despite the challenge in the recovery of the each ejecta parameter, the analytic model still recovers the total ejecta mass relatively well.
This is the case for variety of the merger models as well as viewing angles.
However, the estimate of the total mass is affected by the specific heating rate: this effect is significant for high $Y_e$ cases, where the specific heating rate is lower. Therefore, the uncertainty in the heating rate becomes a significant source of systematic error \citep{Wanajo14, Wu16, Lippuner17, kawaguchi21}.

In this study, we use the complete set of light curves from radiative transfer simulations as mock observational data. However, in actual observations, it is often challenging to obtain a complete light curve.
We demonstrate that missing data can lead to an overestimate in the total mass up to by a factor of about two, as compared with the case with complete data. 
Our experiments show that only optical data are insufficient, i.e., 
infrared observations near the peak of the light curves are crucial to reliably estimate the total ejecta mass.

\begin{acknowledgements}
This work was supported by JST FOREST Program (Grant Number JPMJFR212Y) and the Grant-in-Aid for Scientific Research from JSPS (21H04997, 23H00127, 23H04891, 23H04894, 23H04900, 23H05432).
We thank Hamid Hamidani for valuable discussion.
\end{acknowledgements}


\appendix
\section{Analytic kilonova model}
\label{sec:model}

We describe an outline of our analytic model for kilonova light curve. Our model is broadly the same as that used by \cite{villar17}, which is based on the prescription by \cite{metzger17}, and implemented in \texttt{MOSFiT} \citep{nicholl17, villar17a, guillochon17}.
Our model uses slightly different approximations in the heating rate and thermalization as described below. Note that we use the parameters $M$ (ejecta mass), $v$ (ejecta velocity), and $\kappa$ (opacity) as characteristic values of the ejected materials.  
For the MCMC parameter estimation, we define the velocity $v$ by using the kinetic energy $E=Mv^2/2$, consistent with the definition used in numerical relativity simulations. 
In the analytical model described below, however, we use the velocity of the outer edge of the ejecta ($R=v_{\rm{max}}t$) for the consistency with literature.
Our model adopts one-zone approximation, i.e., we assume the ejecta with a constant density. 
Under this assumption, the relationship between $v_{\rm{max}}$ and $v$ is simply $v=\sqrt{3/5}v_{\rm{max}}$.

The internal energy of ejecta evolves according to
\begin{equation}
  \frac{dE_{\mathrm{int}}}{dt}=-P\frac{dV}{dt}-L(t)+\dot{Q}(t)
  \label{eq:E-evolution}
\end{equation}
where the first term in the right hand side accounts for the energy loss due to adiabatic expansion in the radiation-dominated ejecta, the second term for the radiative losses, and the third term for energy deposition rate by radioactive decays. The first term accounts for the loss due to adiabatic expansion in the radiation-dominated ejecta. It can be described as $-PdV/dt = - E_{\rm{int}}/t$, with the dynamical timescale $t=R/v_\mathrm{max}$, where $R$ represents ejecta radius. 

The second term in Eq.(\ref{eq:E-evolution}) accounts for the radiative losses; determined by the energy escaping on the diffusion timescale:
\begin{equation}
  L(t) = \frac{E_{\rm{int}}}{t_{\rm{diff}}}\ ,
\end{equation}
where the diffusion timescale is
\begin{equation}
  t_{\mathrm{diff}}=\frac{3\kappa M}{4\pi cv_{\mathrm{max}}t}\  
\end{equation}
for the ejecta with a homogeneous density.

The third term in Eq.(\ref{eq:E-evolution}) accounts for heating rate. Kilonovae are powered by the decay of a wide range of $r$-process nuclei with different half-lives, leading to a specific heating rate $\dot{q}(t)$ in a power-law form \citep{metzger10, Roberts11}:
\begin{equation}
  \dot{q}(t) \approx 2\times10^{10}\left(\frac{t}{1\ \mathrm{day}}\right)^{-1.3}\ \ \mathrm{erg\ s^{-1}\ g^{-1}}.
  \label{eq:qdot}
\end{equation}
The total deposition rate $\dot{Q}(t)$ is 
\begin{equation}
  \dot{Q}(t) = f(t)M \dot{q}(t),
  \label{eq:heating}
\end{equation}
where $f(t)$ is the thermalization efficiency. 
Note that \citet{villar17} adopted a specific heating rate by \citet{korobkin12}, which also reproduces a 
roughly constant heating rate during the first few seconds.
We approximate the specific heating rate by a simple power law as the heating in the first few seconds does not have any impact to kilonova emission at a timescale of days.
Radioactive heating occurs through $\beta$-decays, $\alpha$-decays, and fission. The thermalization efficiency $f(t)$ depends on how these decay products share their energies with the thermal plasma. Since $\beta$-particles and $\gamma$-rays from $\beta$-decay dominates the energy deposition in a broad range of $Y_{\rm e}$, we approximate the total thermalization efficiency using their typical fractions as
\begin{equation}
  f(t) \simeq 0.45f_\gamma (t) + 0.2f_e (t)
\end{equation}
where $f_{\gamma}(t)$ and $f_e(t)$ are the thermalization efficiency for $\gamma$-rays and $\beta$-particles, respectively. 
We evaluate the thermalization timescales by using the global ejecta parameter ($M$ and $v$) as in \citet{rosswog17}, which is based on the evaluation by \citet{barnes16}.
This is another difference from the assumptions by \citet{villar17}, where tabulated thermalization efficiency by \citet{barnes16} was adopted with an interpolation. We adopt a simple one-zone prescription to be free from the assumption of the density structure in \citet{barnes16}.

For the spectral energy distribution, we assume a blackbody radiation with a temperature determined by its bolometric luminosity and a radius using the Stefan-Boltzmann law.
We also assume that the photospheric radius equals to ejecta radius ($R_\mathrm{phot} = R$). Based on these assumptions, the forms of photospheric temperature and radius are:
\begin{equation}
  T = \mathrm{max}\left[\left(\frac{L(t)}{4\pi \sigma_{\rm{SB}} v_\mathrm{max}^2 t^2}\right)^{1/4}, \ T_{\mathrm{c}}\right]
\end{equation}
and
\begin{equation}
  R_{\mathrm{phot}}=
  \left\{
    \begin{aligned}
      &v_\mathrm{max}t
      &\left(\frac{L(t)}{4\pi \sigma_{\rm{SB}} v_\mathrm{max}^2 t^2}\right)^{1/4}>T_{\mathrm{c}}\\
      &\left(\frac{L(t)}{4\pi \sigma_{\rm{SB}} T_{\mathrm{c}}^4}\right)^{1/2}
      &\left(\frac{L(t)}{4\pi \sigma_{\rm{SB}} v_\mathrm{max}^2 t^2}\right)^{1/4}\leq T_{\mathrm{c}}
    \end{aligned}
  \right.
\end{equation}
where $T_\mathrm{c}$ is a temperature floor, which phenomenologically represents a critical temperature that the ejecta cools to. 
Although it is not obvious whether the ejecta from a kilonova would exhibit this effect, we have aligned our model with that by \cite{villar17} to ensure consistency.

Our model includes two components: ``blue'' and ``red" components. 
The total flux is assumed to be a simple sum of the fluxes from the red and blue components ($F_\mathrm{tot} = F_\mathrm{red} + F_\mathrm{blue}$).
Each component is characterized by four parameters: mass ($M$), velocity ($v$, which is defined as $E=Mv^2/2$), opacity ($\kappa$), and temperature floor ($T_c$). 

\section{Parameter estimation for GW170817/AT2017gfo}
\label{sec:obsfit}

\begin{figure}[h]
  \begin{center}
    \begin{tabular}{cc}
    \hspace{-1.0cm}
      \includegraphics[width=10cm]{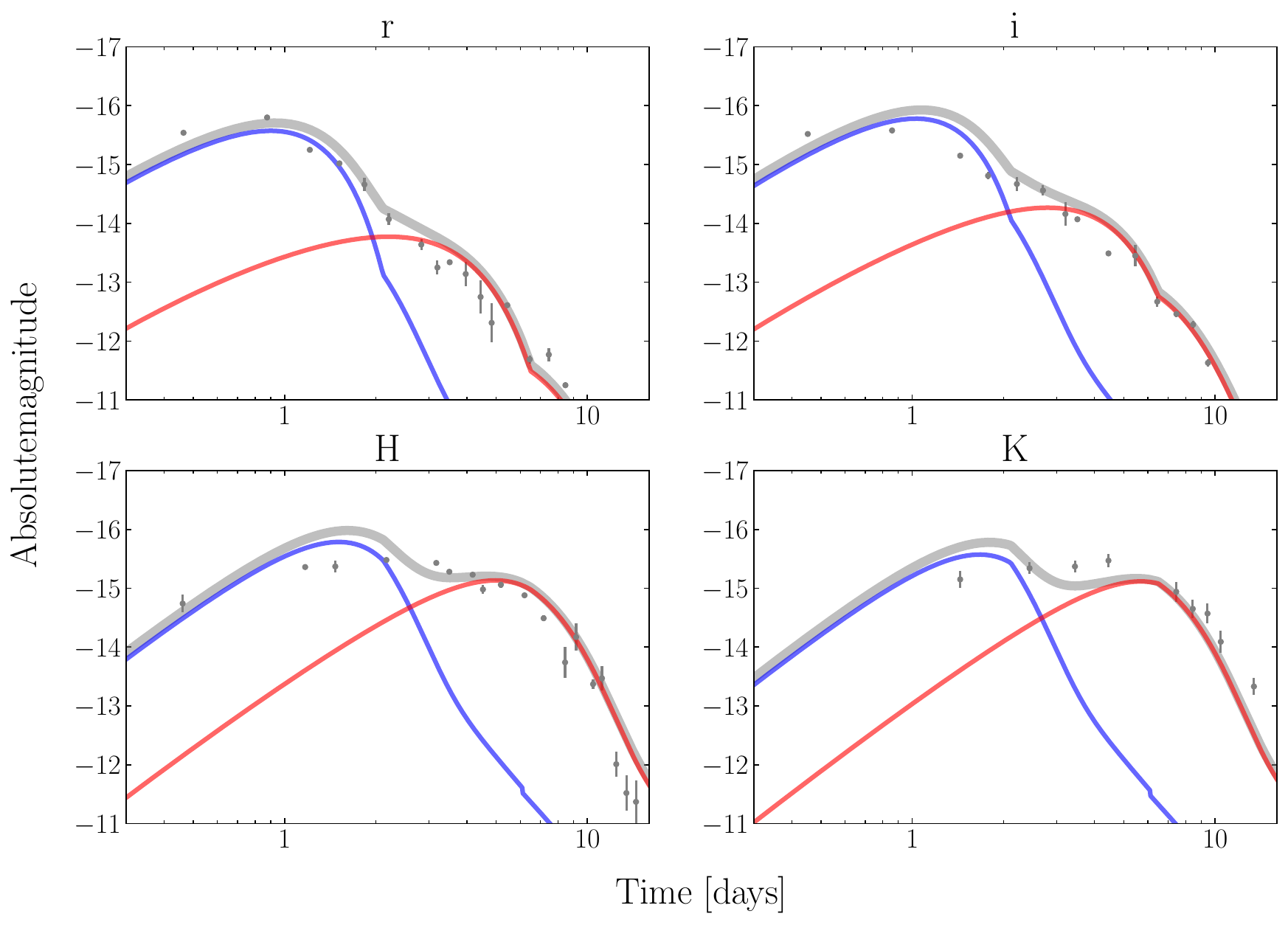}& 
      \includegraphics[width=6cm]{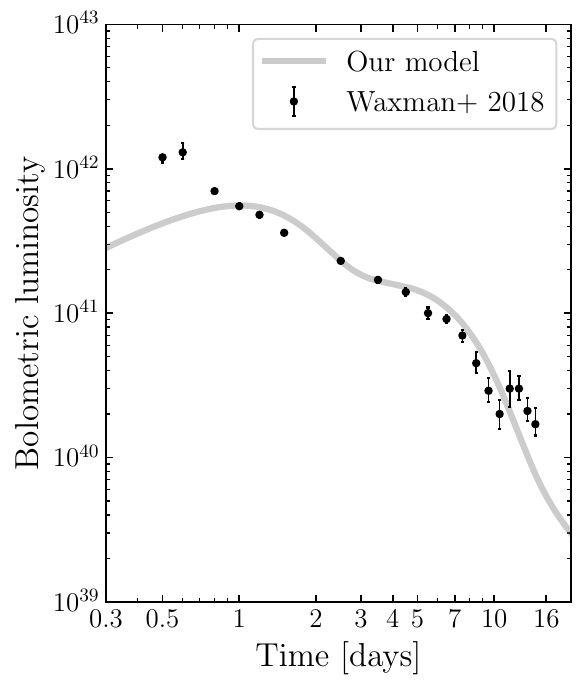}
    \end{tabular}
    \caption{(Left) Individual band light curves of GW170817/AT2017gfo taken from \cite{villar17} (gray circles), the two-component best-fit model (gray lines), and the blue and red components in the model (blue and red lines). (Right) The bolometric luminosity of the best-fit model (line) with that of GW170817/AT2017gfo (points, \citealt{waxman18}).
  }
\label{fig:obs_lc}
\end{center}
\end{figure}

To test the validity of our method, we fit the the observational data of GW170817/AT2017gfo \citep{villar17}. The model realizations with the highest likelihood (“best-fit”) 
show a good agreement with the observational data (left panels of Figure \ref{fig:obs_lc}) as in the two-component model by \cite{villar17}. The bolometric luminosity also matches well with that of GW170187/AT2017gfo (\citealt{waxman18}, right panel of Figure \ref{fig:obs_lc}). As summarized in Table \ref{tab:result-GW170817}, each inferred parameter also shows the same relationship with those of the red and blue components in \cite{villar17} (see Figure \ref{fig:obs_Mv}). The estimated values of the mass for each component are somewhat smaller than those in \cite{villar17}. 
This difference arises from slightly different approximations in the thermalization efficiency(see above).
Also, the estimated total mass is smaller than that in \citet{hotokezaka20} as they adopted different prescription of the thermalization efficiency.
Nevertheless, these differences only affects the overall scale of the estimated masses, and do not affect our conclusions on the physical meaning of blue/red components.

\begin{table*}[h]
  \begin{center}
  \caption{Inferred parameters for GW170817}
  \label{tab:result-GW170817}
  \hspace{-2.5cm}
    \begin{tabular}{ccccccccc}
      \hline \hline
      $M^{\rm{blue}}$ & $v^{\rm{blue}}$ & $\kappa^{\rm{blue}}$ & $T_c^{\rm{blue}}$ & $M^{\rm{red}}$ & $v^{\rm{red}}$ & $\kappa^{\rm{red}}$ & $T_c^{\rm{red}}$ & $\sigma$ \\ \hline
      $0.009_{-0.001}^{+0.001}$ & $0.27_{-0.02}^{+0.03}$ & (0.5) & $2885_{-1561}^{+973}$ & $0.016_{-0.001}^{+0.001}$ & $0.10_{-0.01}^{+0.01}$ & $2.0_{-0.2}^{+0.3}$ & $2380_{-118}^{+97}$ & $0.315_{-0.028}^{+0.033}$\\
      \hline
    \end{tabular}
  \end{center}
\end{table*}

\bibliography{kitamura.bib}
\bibliographystyle{aasjournal}



\end{document}